\documentclass{PoS}

\title{Time-resolved multiwavelength observations of the blazar
  VER~J0521+211 from radio to gamma-ray energies}

\ShortTitle{Time-resolved MWL observations of the blazar VER~J0521+211}

\author{\speaker{H. Prokoph}$^a$ for the
  VERITAS~Collaboration\thanks{veritas.sao.arizona.edu} \\ 
  \newline 
  \llap{$^a$}Department of Physics and Electrical Engineering, Linnaeus
  University, 351 95 V\"axj\"o, Sweden \\ 
  E-mail: \email{heike.prokoph@lnu.se} \\

}

\author{P. Da Vela$^b$ and C. Schultz$^c$ for the
  MAGIC~Collaboration\footnote{magic.mpp.mpg.de} \\ 
  \newline
  \llap{$^b$}Universit\`a  di Siena and INFN Pisa, I-53100
  Siena, Italy \\
  \llap{$^c$}Universit\`a di Padova and INFN, I-35131 Padova, Italy 
}

\abstract{VER~J0521+211 (RGB~J0521.8+2112) is one of the brightest and
  most powerful blazars detected in the TeV gamma-ray regime. It is
  located at a redshift of $z=0.108$ and since its discovery in 2009,
  VER~J0521+211 has exhibited an average TeV flux exceeding 0.1 times
  that of the Crab Nebula, corresponding to an isotropic luminosity of
  $3\times10^{44}$\,erg\,s$^{-1}$. 
  We present data from a comprehensive multiwavelength
  campaign on this object extending between November 2012 and February
  2014, including single-dish radio observations, optical photometry
  and polarimetry, UV, X-ray, GeV and TeV gamma-ray data (VERITAS,
  MAGIC). Significant flux variability was observed at all
  wavelengths, including a long-lasting high state at gamma-ray
  energies in Fall 2013. Nightly-resolved spectra at X-ray and TeV
  energies are be presented, and emission mechanisms explaining the
  observed flux and spectral variability are discussed.} 

\FullConference{The 34th International Cosmic Ray Conference,\\
		30 July- 6 August, 2015\\
		The Hague, The Netherlands}

\def\degree{^{\circ}}
\def\ver{VER~J0521+211}
\def\rgb{RGB~J0521.8+2112}

\def\apj{ApJ}

\newcommand{\reffig}[1]{Figure~\ref{#1}}  % figure references
\begin{document}
%%%%%%%%%%%%%%%%%%%%%%%%%%%%%%%%%%%%%%%%%%%%%%%%%%%%%%%%%%%%%%%%%%%%%%%%

\section{Introduction}

%%% HP (v2->v3): small changes...
% less numbers from the discovery paper to reduce confusion for the reader

\ver, spatially associated with the radio and X-ray source \rgb, was
discovered in 2009 as TeV emitter by VERITAS \cite{0521-atel1} and is
now unambiguously classified as a BL~Lac-type blazar \cite{0521}. 
As usually observed in blazars, \ver\ is highly variable at all
wavelengths. With an average integral flux above 200~GeV of
about $>0.1$ Crab\footnote{The gamma-ray flux of the Crab Nebula
  used here is $F(E>200\,\mathrm{GeV}) = 2.1 \times 10^{-10}
  \mathrm{cm}^{-2} \mathrm{s}^{-1}$ \cite{crab-hillas}.}, 
\ver\ is among the brightest known TeV blazars. Located at a redshift of
$z=0.108$ \cite{shaw}, this flux corresponds to a gamma-ray
luminosity of about $3 \times 10^{44}\mathrm{erg}\,\mathrm{s}^{-1}$ 
and is thus significantly brighter than the classical northern
high-synchrotron-frequency-peaked BL Lacs (HBLs) such as Mrk~421, Mrk~501
and 1ES~1959+650. \ver\ itself shows a synchrotron peak in the optical
regime, suggesting a classification as intermediate-frequency-peaked
BL Lac (IBL). However, during the TeV high state in November 2009
\cite{0521-atel2}, \ver\ showed more HBL-like properties, especially
given the observed spectral hardening in the X-ray regime with
increased flux. The observed variability and its bright TeV flux
motivated further multiwavelength observations of \ver\ in
order to extend the detailed time-resolved spectral modeling available
for nearby HBLs \cite{sed-1959,sed-421,sed-501} to a more luminous blazar which are reported here. 

%In the TeV gamma-ray regime, no significant spectral
%variability was found and the time-averaged energy spectrum above
%200~GeV can be well described by a power law with a photon index of 
%$\Gamma = 3.44\pm0.20$. 

%%%%%%%%%%%%%%%%%%%%%%%%%%%%%%%%%%%%%%%%%%%%%%%%%%%%%%%%%%%%%%%%%%%%%%%%
\section{VERITAS}

\begin{figure}[ht]
\centering
\includegraphics[width=0.49\textwidth]{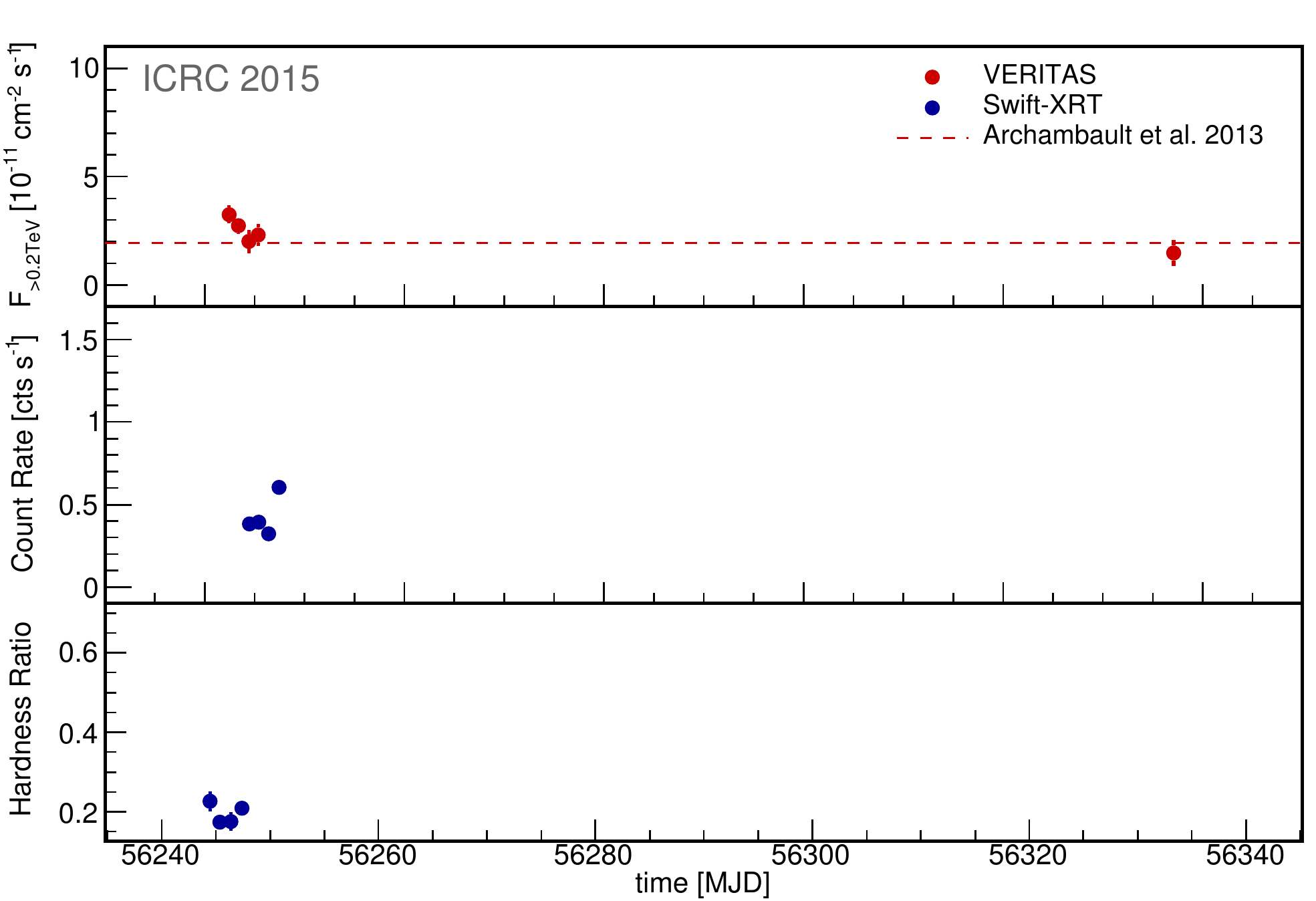}
\includegraphics[width=0.49\textwidth]{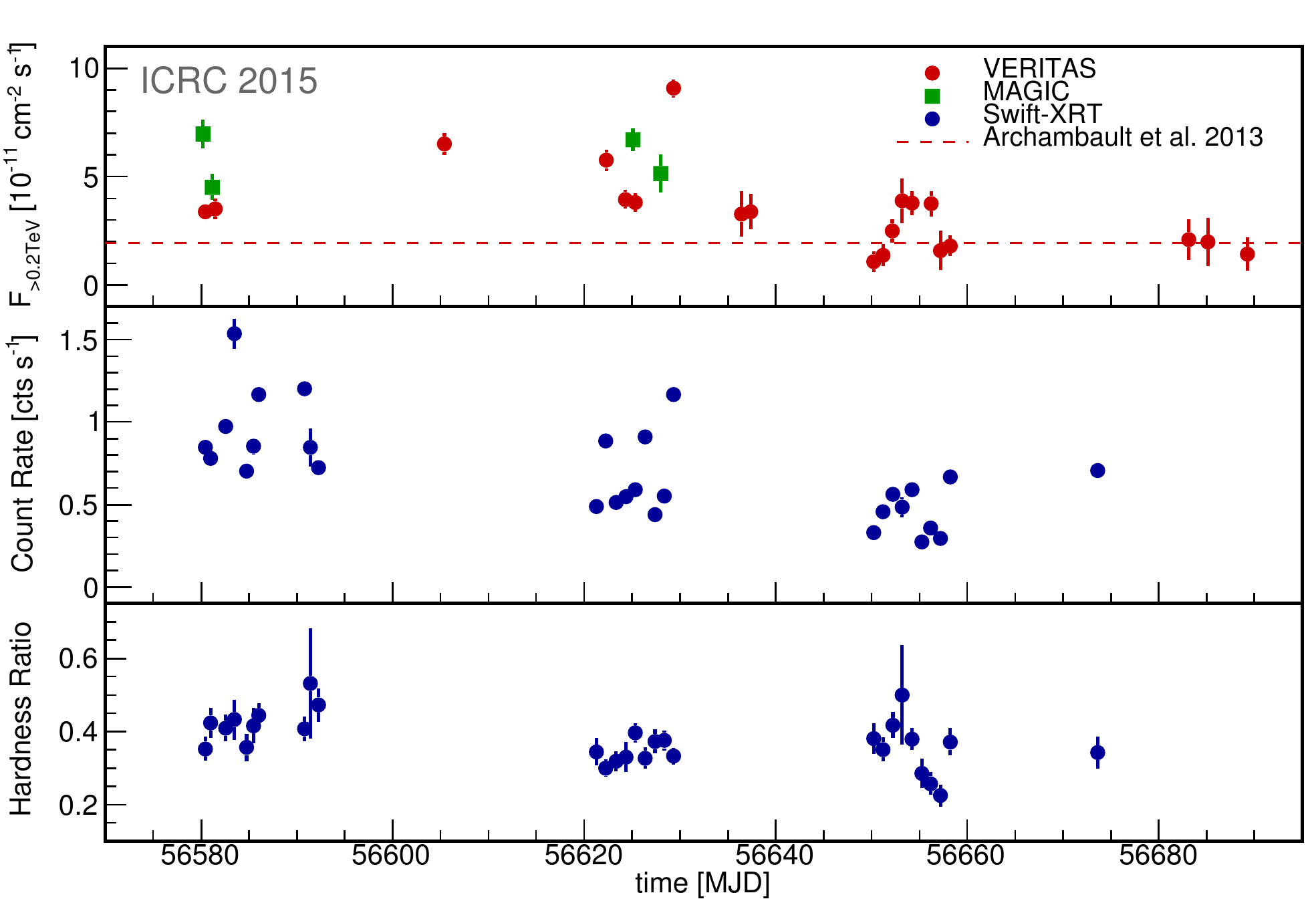}
\caption{Evolution of the TeV flux, X-ray flux, and X-ray hardness
  ratio as a function of time during the 2012 ({\it left}) and 2013
  seasons ({\it right}). It should be noted that the MAGIC and VERITAS
  observations in 2013 were not taken simultanously.}
\label{fig:lightcurve}
\end{figure}

%%% VERITAS instrument description
VERITAS is an array of four 12\,m-diameter imaging atmospheric Cherenkov
telescopes (IACTs), located in southern Arizona, USA. It is sensitive
to gamma-ray energies between 85\,GeV and $>$30\,TeV and is capable
to detect a source with 0.01\,{Crab} flux in under $\sim 25$\,{hours} of
observations. 
%%% VERITAS data set 
VERITAS observations of \ver\ reported here were taken in two seasons:
the 2012 season comprising data taken between 2012 Nov 11 and 2013 Feb
14; and the 2013 season (2013 Oct 15 to 2014 Feb 01). 
All data were taken in {\it wobble mode} for which the source is
observed with an offset of $0.5^{\circ}$ with respect to the center of
the instrument's field of view to allow for simultaneous background
measurements \cite{fomin94}. After quality selection, the dataset
comprised 5.0\,h live time in the 2012 season and 16.5\,h in the 2013 season.
%%% VERITAS data analysis 
VERITAS data analysis followed the procedure outlined in \cite{vegas}.
Cherenkov light from air showers initiated by gamma rays and
cosmic rays triggers the readout of PMT
signals, which are then calibrated and used to reconstruct an image of
the shower in the focal plane. Individual telescope images are
parameterized by moment analysis, and
geometrical reconstruction is used to estimate the arrival direction, $\theta$,
of the primary particle with respect to the source location.
%with pre-defined cuts, designed to provide optimum sensitivity to a
%{\red moderately strong point-like gamma-ray source (0.05\,{Crab})
%  with a Crab-like differential power-law photon index of $\sim 2.5$}. 
The signal was extracted using a reflected region background model
with an ON region of 0.17$\degree$ radius centered on
the position of \ver\ (R.A. = $05^{\rm h}21^{\rm m}45^{\rm s}$,
Decl. = $+21^{\circ}12^{\prime}51.\!^{\prime\prime}4$, J2000). 
%%% VERITAS results
The strength of the gamma-ray signal is evaluated using Eq.~17 in
\cite{lima} and was found to be $19.0\,\sigma$ in 2012 and
$60.3\,\sigma$ in 2013. The average flux levels in the two seasons are
$F_{>0.2\,\mathrm{TeV}}^{`12}=2.4 \pm 0.2$ and
$F_{>0.2\,\mathrm{TeV}}^{`13}=4.2 \pm 0.1$ in units of
$10^{-11}\,\mathrm{cm}^{-2}\,\mathrm{s}^{-1}$. For comparison, the
time averaged flux in \cite{0521} was $1.9 \pm 0.1$ in the same units
(as shown in \reffig{fig:lightcurve}).

%%%%%%%%%%%%%%%%%%%%%%%%%%%%%%%%%%%%%%%%%%%%%%%%%%%%%%%%%%%%%%%%%%%%%%%%
\section{MAGIC}

\begin{figure}[htp]
\centering
\includegraphics[width=4.in]{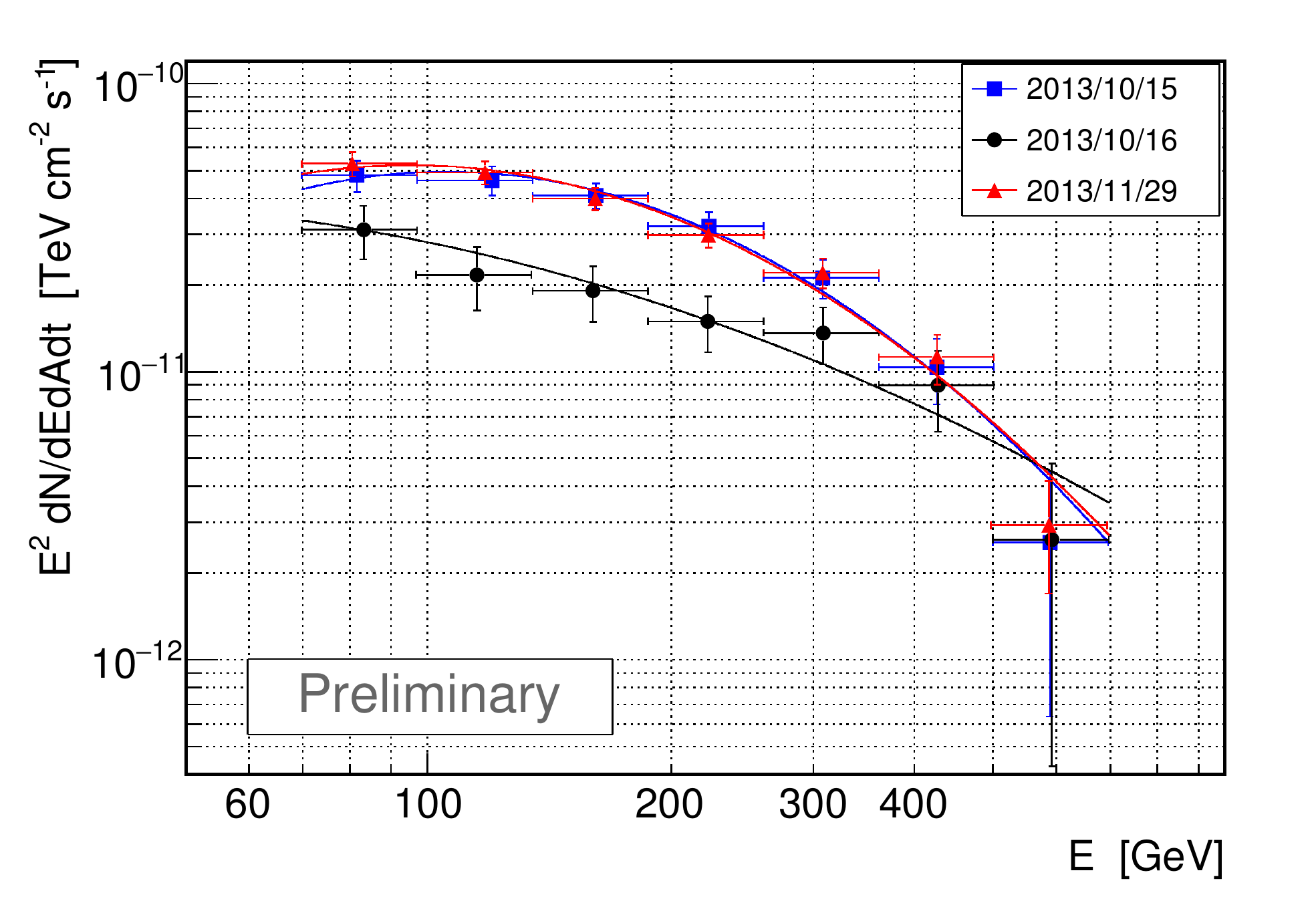}
\caption{Spectral energy distribution of \ver\ as obtained by MAGIC
  during three single nights in 2013.}
\label{fig:magic-spec}
\end{figure}

%%% magic intro
MAGIC is a stereoscopic system of two 17~m-diameter IACTs, located at
La Palma, Canary Islands. It is sensitive to gamma-ray energies above
$\sim50$~GeV (in normal trigger mode) and underwent a series of
upgrades during summer 2011 and 2012 \cite{magic-upgrades}. In
stereoscopic mode, the integral sensitivity for sources with Crab
Nebula-like spectra above 220 GeV is $(0.66\pm0.03)$\% of Crab Nebula
flux in 50\,h of observations (more details about the MAGIC
performance are given in \cite{MAGIC_performance}). 
%%% magic data set
MAGIC observations of \ver\ were taken between 2013 Oct~15 and Dec~2 and
were triggered by the high state in optical and high-energy gamma rays
reported by KVA (E. Lindfors, private communication) and {\it Fermi}-LAT \cite{atel5472},
respectively. Observations covering a zenith angle range from 
7$\degree$ to 26$\degree$ were performed in {\it wobble mode} with an
offset of $0.4^{\circ}$. After quality selection and dead time correction,
the total effective observation time was $\sim4.5$\,hours. 
%%% magic analysis decription 
The data were analyzed using the MAGIC analysis and reconstruction 
software package \cite{zanin}, which has been adapted to stereoscopic
observations \cite{magic_stereo}. 
%For the reconstruction of the shower arrival direction the random
%forest regression method (RF DISP method; Aleksi{\'c} et al. 2010)
%with the implementation of stereoscopic parameters such as the impact
%distance of the shower on the ground was used (Lombardi et
%al. 2011). The gamma/hadron separation was performed by using the
%random forest method (Albert et al. 2008) which is based on both
%individual image parameters from each telescope and stereoscopic
%information such as the shower impact point and the shower height maximum. 
%%% magic results
The signal was extracted from an ON region of 0.11$\degree$ radius centered on the position of \ver. 
The analysis of the overall data set yielded a strong signal of 
30.5\,$\sigma$ significance \cite{lima}, while the significances of
the signal from the 
individual nights are 18.6\,$\sigma$, 9.7\,$\sigma$, 21.9\,$\sigma$
and 6.2\,$\sigma$ for Oct 15, 16, Nov 29 and Dec 2, respectively. Given the significance of the signal, night-wise spectra
could be obtained above the energy threshold of 65\,GeV for the
observations in October and November and are shown in \reffig{fig:magic-spec}. 

\begin{figure}[htp]
\centering
\includegraphics[width=4.5in]{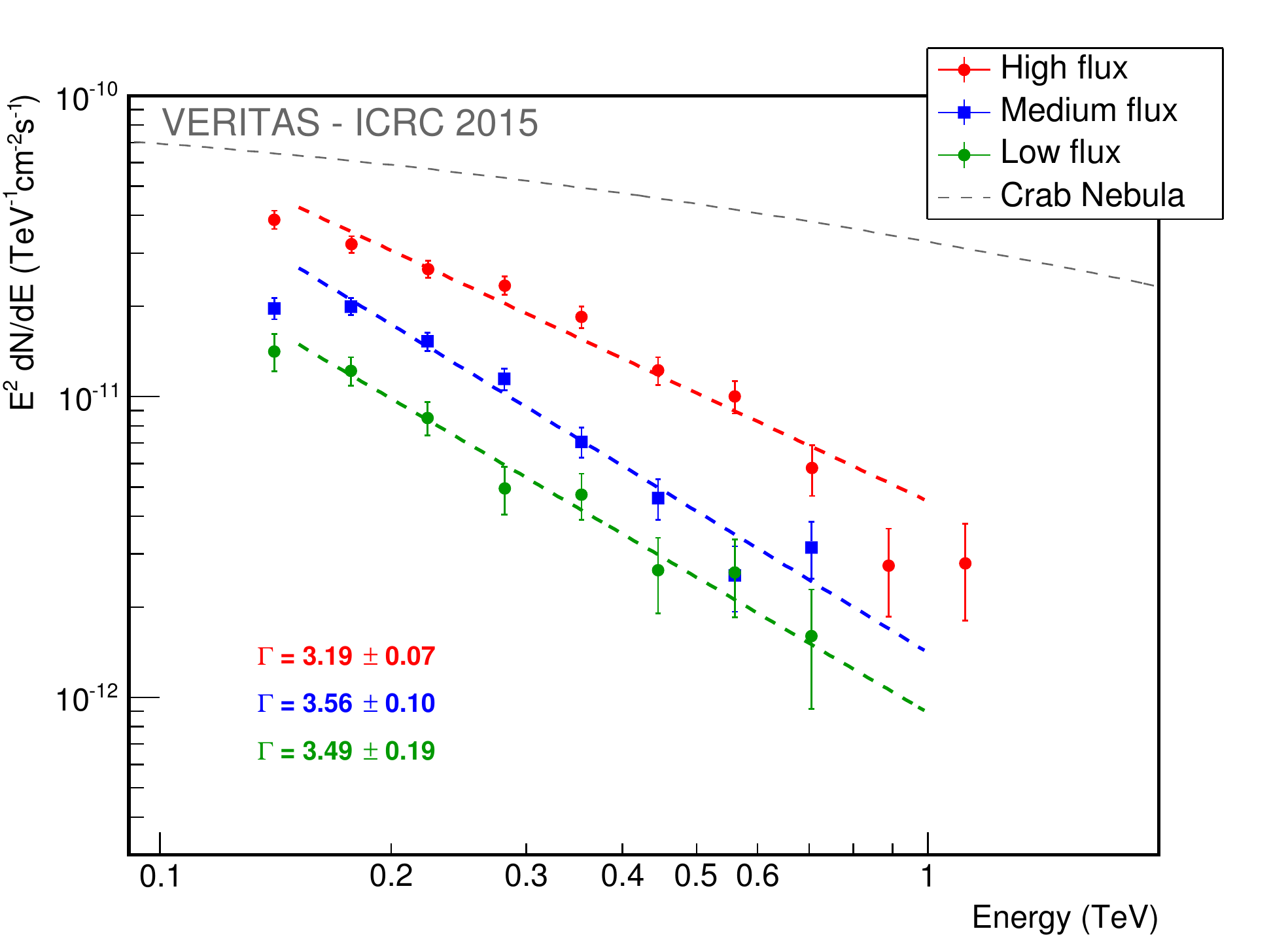}
\caption{Comparison of the spectral energy distributions measured with VERITAS at different flux states. Spectra have been obtained by stacking observations with nightly fluxes $F_{>0.2\,\mathrm{TeV}} < 3 \times 10^{-11}\,\mathrm{cm}^{-2}\,\mathrm{s}^{-1}$ (\emph{low} flux), $F_{>0.2\,\mathrm{TeV}} > 4 \times 10^{-11}\,\mathrm{cm}^{-2}\,\mathrm{s}^{-1}$ (\emph{high} flux), and a \emph{medium} flux bin defined between these two values. }
\label{fig:spectrum}
\end{figure}

%%%%%%%%%%%%%%%%%%%%%%%%%%%%%%%%%%%%%%%%%%%%%%%%%%%%%%%%%%%%%%%%%%%%%%%%
%\section{Fermi/LAT}

%%%%%%%%%%%%%%%%%%%%%%%%%%%%%%%%%%%%%%%%%%%%%%%%%%%%%%%%%%%%%%%%%%%%%%%%
\section{X-ray} 
The {\it Swift} X-ray telescope (XRT \cite{{burrows2005}}) monitored \ver\ in the X-ray band (0.3-10\,keV) for most of the nights of the campaign. Data were processed using the most recent versions of the standard Swift tools: Swift Software version 3.9, FTOOLS version 6.12 and XSPEC version 12.7.1. Light curves are generated using {\tt xrtgrblc} version 1.6. More details on the monitoring program and the analysis procedure can be found in \cite{stroh}.

%%%%%%%%%%%%%%%%%%%%%%%%%%%%%%%%%%%%%%%%%%%%%%%%%%%%%%%%%%%%%%%%%%%%%%%%
\section{Flux and spectral variability}
The TeV fluxes measured in 2012 and 2013 are higher than those
reported during the discovery of the source \cite{0521}, with the
average flux in 2013 being a factor of $2$ brighter than that
of 2009. The nightly flux (Figure~\ref{fig:lightcurve}) shows
significant variability down to timescales of 1 day. 

To test for spectral variability, we have produced gamma-ray spectra for
individual observing nights (Figure~\ref{fig:magic-spec}) and by
stacking observations at different flux levels
(Figure~\ref{fig:spectrum}). At low and medium flux levels we find
gamma-ray spectra largely compatible with a power-law with index
$\Gamma = 3.44 \pm 0.20$ that was reported in previous observations
\cite{0521}. Co-adding the spectra from the three nights with higher
fluxes we measure $\Gamma=3.10 \pm 0.07$, constituting evidence for
a mild hardening of the gamma-ray spectrum when the emission is
brighter. In addition, the lower threshold of the MAGIC observations
(Figure~\ref{fig:magic-spec}) helps to reveal some evidence of
spectral curvature, suggesting a peak of the inverse-Compton emission
component at energies of $\sim 100$\,GeV. 

\begin{figure}[htb]
\centering
\includegraphics[width=4.in]{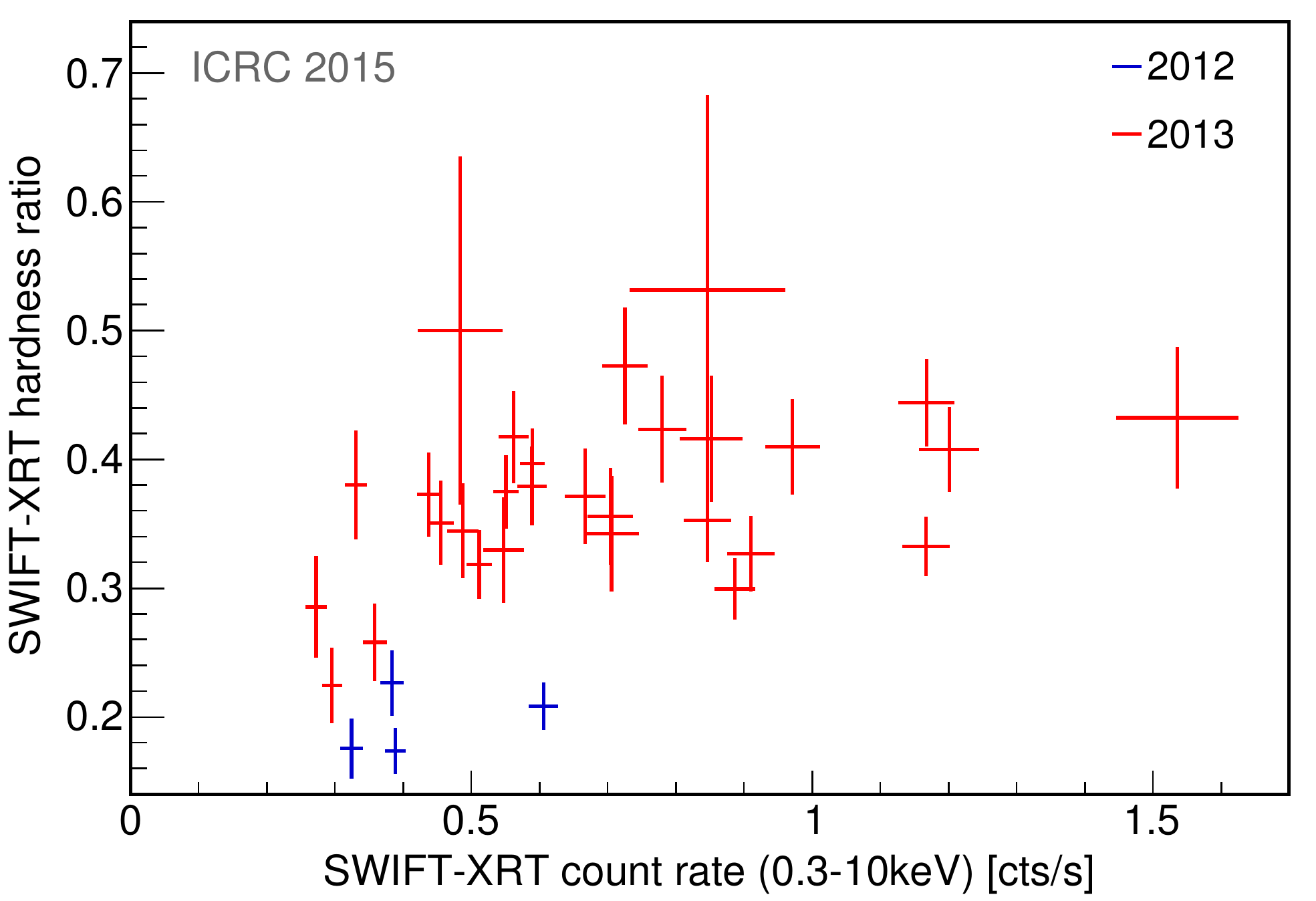}
\caption{X-ray hardness ratio as a function of X-ray flux measured by
  {\it Swift}-XRT. The hardness ratio is defined as the rate in the
  2-10\,keV band divided by the rate in the 0.3-2\,keV band.   }
\label{x}
\end{figure}

\begin{figure}[htb]
\centering
\includegraphics[width=4.in]{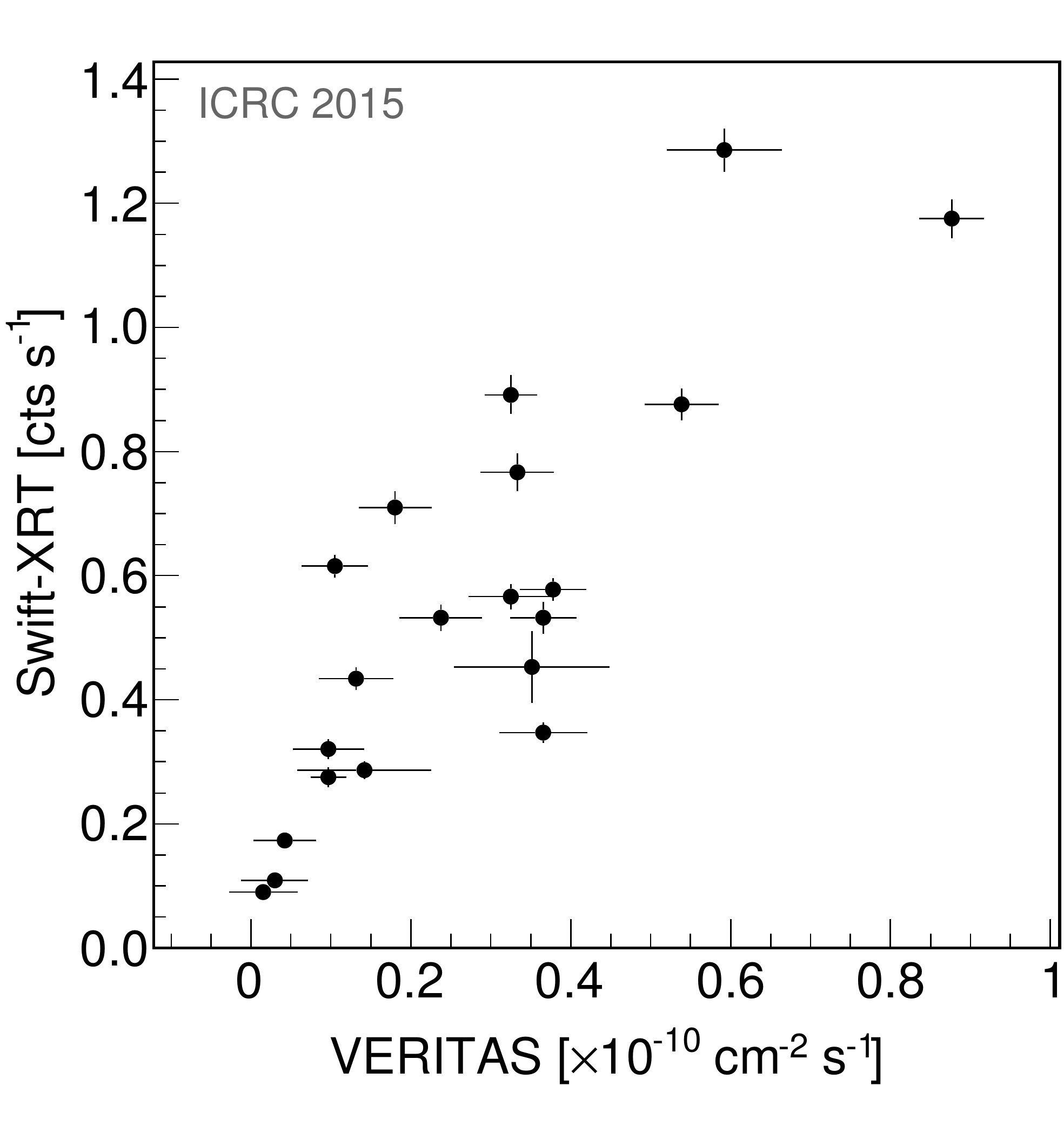}
\caption{X-ray flux as a function of the TeV flux for nights with simultaneous observations in both bands.}
\label{tev_x}
\end{figure}

Similar levels of 1-day-scale flux variability can be seen in the
X-ray band (Figure~\ref{fig:lightcurve}). The evolution of the hardness
ratio in the X-ray band (Figure~\ref{x}) suggests only small changes
of the X-ray spectral shape at different flux levels. Data from the nights with simultaneous observations by VERITAS and {\it Swift}-XRT (Figure~\ref{tev_x}) show a clear correlation between the measured fluxes in both bands.

%%%%%%%%%%%%%%%%%%%%%%%%%%%%%%%%%%%%%%%%%%%%%%%%%%%%%%%%%%%%%%%%%%%%%%%%
\section{Discussion}
The data presented in these proceedings constitute the first extensive multi-wavelength campaign on \ver. Significant changes in the X-ray and TeV fluxes from the source were seen in timescales of $\sim 1$ day, with an amplitude of approximately a factor of $5$ in both bands. Spectral variability in the TeV band is first reported here, displaying a mild hardening of the spectrum at higher flux levels. The spectral variability is however much less significant than typically found in HBLs. 

The observed correlation between the fluxes in the X-ray and TeV bands suggests a connection between the synchrotron and the high-energy spectral components of the emission, as expected if the high-energy emission is dominated by synchrotron self-Compton. However, a contribution from inverse-Compton emission from electrons and positrons interacting with other radiation fields external to the jet cannot be excluded. 

Compared to similar studies conducted on TeV HBLs (Mrk~421, Mrk~501, 1ES~1959+650), the level of spectral variability in the TeV regime is small given the high-significance of the detections and the TeV flux changing by a factor of $\sim 5$. This small level of spectral variability and a softer spectral index seem characteristic of IBLs, where electrons and positrons with energies beyond the TeV scale will be less dominant than in HBLs given the higher rate of inverse-Compton cooling. 

\vspace{1cm} 

\emph{VERITAS is supported by grants from the U.S. Department of Energy Office of Science, the U.S. National Science Foundation and the Smithsonian Institution, and by NSERC in Canada. We acknowledge the excellent work of the technical support staff at the Fred Lawrence Whipple Observatory and at the collaborating institutions in the construction and operation of the instrument. The VERITAS Collaboration is grateful to Trevor Weekes for his seminal contributions and leadership in the field of VHE gamma-ray astrophysics, which made this study possible. }

\emph{MAGIC would like to thank the Instituto de Astrof\'{\i}sica de Canarias for the excellent working conditions at the Observatorio del Roque de los Muchachos in La Palma. The financial support of the German BMBF and MPG, the Italian INFN and INAF, the Swiss National Fund SNF, the ERDF under the Spanish MINECO (FPA2012-39502), and the Japanese JSPS and MEXT is gratefully acknowledged. This work was also supported by the Centro de Excelencia Severo Ochoa SEV-2012-0234, CPAN CSD2007-00042, and MultiDark CSD2009-00064 projects of the Spanish Consolider-Ingenio 2010 programme, by grant 268740 of the Academy of Finland, by the Croatian Science Foundation (HrZZ) Project 09/176 and the University of Rijeka Project 13.12.1.3.02, by the DFG Collaborative Research Centers SFB823/C4 and SFB876/C3, and by the Polish MNiSzW grant 745/N-HESS-MAGIC/2010/0. }

%%%%%%%%%%%%%%%%%%%%%%%%%%%%%%%%%%%%%%%%%%%%%%%%%%%%%%%%%%%%%%%%%%%%%%%%

\end{document}